\documentclass[aps,pra,preprint,groupedaddress,showpacs,showkeys]{revtex4}

\usepackage {graphicx}
\usepackage {tabularx}
\usepackage{color}
\usepackage {ulem}
\usepackage{amsmath}
\usepackage{enumerate}

\begin{document}
\def\vector#1{\mbox{\boldmath $#1$}}
\def\vectorsc#1{\mbox{\boldmath \scriptsize{$#1$}}}

\title{Calculations of Magnetic properties of metals through 
the magnetic-field-containing relativistic tight-binding 
approximation method}



\author{Masahiko Higuchi}
\affiliation{Department of Physics, Faculty of Science, Shinshu University, 
Matsumoto 390-8621, Japan}

\author{Katsuhiko Higuchi}
\affiliation{Graduate School of Advanced Sciences of Matter, 
Hiroshima University, Higashi-Hiroshima 739-8527, Japan}


\date{\today}

\begin{abstract}
Magnetic properties of metals are investigated 
through electronic structure calculations based on the recently-proposed 
magnetic-field-containing relativistic tight-binding approximation (MFRTB) 
method [Phys. Rev. B \textbf{91}, 075122 (2015)]. It is found that electronic energy 
bands for the metal immersed in the uniform magnetic field have a cluster 
structure in which multiple energy bands lie within a small energy width. 
Each cluster corresponds to the energy level that is derived on the basis of 
the semiclassical approximation. While the cluster is responsible for the de 
Haas-van Alphen (dHvA) oscillations, constituent energy bands of the cluster 
cause additional oscillation peaks of the magnetization. Also, the energy 
width of the cluster leads to the reduction of the amplitude of the dHvA 
oscillations, which can be observed as the pseudo Dingle temperature and/or 
the overestimation of the curvature of the Fermi surface. 
\end{abstract}


\maketitle


\section{\label{secI}Introduction}
Measurements of de Haas-van Alphen (dHvA) oscillations 
in metals \cite{1,2,3,4} are 
widely used in investigating the shape of the Fermi surface, cyclotron 
effective mass and relaxation time for scattering of 
electrons \cite{5,6,7}. In order to describe the dHvA oscillations, we need 
electronic states of metals immersed in the uniform magnetic field. For this 
aim, there are conventionally two kinds of methods. One is based on the 
effective Hamiltonian that is obtained by replacing the rest mass of 
electrons with the effective mass in the Hamiltonian for a free electron 
immersed in the uniform magnetic field \cite{5,8,9}. The effects of the periodic 
potential are taken into account via the effective mass. Although quantized 
energy levels (so-called Landau levels) are obtained within this 
approximation, we cannot explain the dependence of the dHvA oscillations on 
the direction of the magnetic field \cite{9}. This is due to an oversimplified 
argument such that the characteristics of individual metals are taken 
into consideration only through the effective mass \cite{9}. 

Another method to describe electronic states of metals immersed in the 
uniform magnetic field is based on the semiclassical approximation 
\cite{4,5,8,9}. Hereafter we shall call this method as the ``semiclassical 
approach''. In the semiclassical approach, instead of directly solving the 
Schr\"{o}dinger or Dirac equation, both the Bohr-Sommerfeld quantization 
rule that is obtained within the semiclassical approximation and the 
equation of motion for a Bloch electron in the magnetic field are used in 
order to get quantized energy levels (semiclassical energy levels) 
\cite{4,5,8,9}. This method leads to the usual description for the dHvA 
oscillations such that every time one semiclassical energy level crosses the 
Fermi energy with increasing the magnetic field, one oscillation of the 
magnetization is produced \cite{5,8,9}. The Lifshitz-Kosevich (LK) formula \cite{10} 
is derived by means of semiclassical energy levels, and is commonly employed 
in analyzing the dHvA oscillations \cite{5,6,7,8,9,10}. On the basis of the LK formula, 
one can evaluate the extremal cross-sectional area of the Fermi surface 
normal to the magnetic field from the oscillation period \cite{5,6,7,8,9,10}. Also, 
according to the LK formula, the temperature and magnetic field dependences 
of the oscillation amplitude give the information on the cyclotron effective 
mass and relaxation time for scattering of electrons, respectively \cite{5,6,7,8,9,10}. 

Recently, we have developed the magnetic-field-containing relativistic 
tight-binding approximation method (MFRTB method) that enables us to 
directly solve the Dirac equation for crystalline materials immersed in the 
uniform magnetic field \cite{11,12}. This method is the first-principles 
calculation method that is applicable to various kinds of realistic 
materials immersed in the uniform magnetic field \cite{11}. 
In the previous work \cite{11}, we have applied this 
method to the crystalline silicon immersed in the magnetic field as the 
first step toward the revealing of the mechanism of the elastic softening 
and its suppression observed in the boron-doped silicon \cite{13,14,15,16}. It is shown 
that the energy band structures have the explicit dependence on the magnetic 
field, and that the recursive energy spectrum which is similar to the 
Hofstadter butterfly diagram \cite{17} is observed. Through this application, 
the MFRTB method is 
illustrated to be useful for revealing the electronic structure of materials 
immersed in the uniform magnetic field \cite{11}. 

Following the above-mentioned application, the MFRTB method is also used to 
describe the dHvA oscillations \cite{12}. It is shown that the dHvA 
oscillations are revisited directly through the MFRTB method  \cite{12}. 
Also, we found 
that the oscillation period of the conventional LK formula is a good 
approximation to that of the MFRTB method in the experimentally available 
magnetic field, while in the high magnetic field it deviates from the period 
of the MFRTB method \cite{12}. However, the detail description of the magnetic 
oscillations of metals through the MFRTB method have not yet been done. 

In this paper, by means of the MFRTB method, we present the detail 
description of magnetic oscillations through the electronic structure of 
metals immersed in the uniform magnetic field. 
Especially, by means of the MFRTB method, we intend to 
investigate unconventional oscillation phenomena that cannot be explained 
by the semiclassical approach.  
For this aim, the MFRTB method is applied to the simple cubic lattice system 
with $s$-electrons that is immersed in the uniform magnetic field \cite{12}. The 
reason why we apply the MFRTB method to this system is that physical 
quantities such as the extremal cross section of the Fermi surface, 
cyclotron effective mass, curvature of the Fermi surface, and so on, can be 
calculated exactly.  
This enables us to investigate unconventional oscillation phenomena 
that cannot be explained by the semiclassical approach, because we can obtain 
rigorous results of the semiclassical approach \cite{12}.

In order to get the full quantum description of the magnetic oscillations, 
we reveal the relation between the semiclassical energy level and the 
electronic structure calculated by the MFRTB method. As shown latter, the 
semiclassical energy level corresponds to the cluster of multiple energy 
bands lying within a small energy width. In other words, the semiclassical 
energy level further splits into multiple energy bands due to the full 
quantum treatment of the MFRTB method. In this paper, we refer to this 
electronic structure as the ``fine energy-level structure''. It is shown 
that this fine energy-level structure becomes obvious with increasing the 
magnetic field, and plays a crucial role for understanding magnetic 
oscillations. For example, it is found that while the conventional dHvA 
oscillations are  
produced by the cluster of energy bands, additional 
oscillation peaks of the magnetization are produced by constituent energy 
bands of the cluster. 

According to the LK formula, the amplitude of the dHvA oscillations depends 
on three quantities, i.e., the cyclotron effective mass, curvature of the 
Fermi surface and relaxation time for scattering of electrons \cite{10}. Since 
the MFRTB method can deal with only zero-temperature systems, it is 
difficult to simultaneously analyze contributions of three quantities to the 
oscillation amplitude. Accordingly, before analyzing the amplitude of the 
dHvA oscillations, we estimate the cyclotron effective mass separately by 
using the density of state (DOS) that is calculated by the MFRTB method. 
Estimated values of the cyclotron effective mass suggest that the 
semiclassical approach gets worse with increasing the magnetic field. 

With use of estimation results of the cyclotron effective mass, the 
oscillation amplitude is analyzed through the MFRTB method. Analysis of the 
oscillation amplitude reveals that the oscillation amplitude is unexpectedly 
reduced in the high magnetic field region, where ``unexpectedly'' means that 
the reduction of the oscillation amplitude cannot be explained by the 
conventional LK formula. The unexpected reduction of the oscillation 
amplitude is caused by the above-mentioned fine energy-level structure. It 
will be shown that this reduction would lead to the observation of the 
``pseudo'' Dingle temperature \cite{18} and/or overestimation of the curvature of 
the Fermi surface even though the relaxation time of electron scattering is 
very long. 

Furthermore, in order to guarantee the validity of calculation results by 
the MFRTB method, the theoretical validity of the MFRTB method is discussed 
in this paper. We investigate the application range of magnitude of the 
magnetic field for the MFRTB method. 
Also, we explain what kind of the boundary condition is imposed on the wave 
function in order to deal with the infinitely large system immersed in a 
uniform magnetic field. 

The organization of this paper is as follows. After a brief explanation of 
the MFRTB method (Sec. II A), we discuss the applicability of the MFRTB 
method to the system immersed in the high magnetic field (Sec. II B). In 
Sec. II C, we explain how to deal with the infinitely large system immersed 
in a uniform magnetic field. In Sec. III, the full quantum description of 
the dHvA oscillations is presented by using the electronic structure 
calculated by the MFTRB method.  In Sec. IV, the appearance, origin 
and observability of additional oscillation peaks are discussed on the 
basis of the detailed investigation of the electronic structure 
calculated by the MFRTB method. In Sec. V, we discuss the limit 
of the semiclassical approach through the estimation of the cyclotron 
effective mass. In Sec. VI, it is shown that the pseudo Dingle temperature 
and/or the overestimation of the curvature of the Fermi surface 
would be observed due to the fine energy-level structure. 
Finally, concluding remarks are presented in Sec. VII. 
%
\section{MFRTB method and its application range}
\label{secII}
In this section, we briefly explain the MFRTB method \cite{11} for the 
convenience of the later discussion. Then, we apply the MFRTB method to the 
simple cubic lattice system with $s$-electrons that is immersed in the 
uniform magnetic field (Sec. II A). Since the effect of the magnetic field 
is treated as the perturbation theory in the MFRTB method, the application 
range of the MFRTB method is discussed before discussing calculation results 
(Sec. II B). In addition, we explain the boundary condition that is used in 
the present calculations so as to treat the infinity large system immersed 
in the uniform magnetic field (Sec. II C). 
%
\subsection{MFRTB method and its application to the simple cubic lattice immersed in the magnetic field}
\label{secII-A}
The Dirac equation for an electron that moves in both the uniform magnetic 
field and periodic potential of the crystal is given by
\begin{equation}
\label{eq1}
\left[ {c{\vector{ \alpha }}\cdot \left\{ {\left. {{\vector{ p}}
+e{\vector{A}}({\vector{r}})} \right\} } \right.+\beta mc^{2}
+\sum\limits_n {\sum\limits_i 
{v_{a_{i} } } ({\vector{r}}-{\vector{R}}_{n} -{\vector{d}}_{i} )} } \right]
\Phi_{{\vectorsc{k}}} ({\vector{r}})=E({\vector{k}})\Phi_{{\vectorsc{k}}} ({\vector{r}}),
\end{equation}
where ${\vector{A}}({\vector{r}})$ and $v_{a_{i} } ({\vector{r}}-{\vector{R}}_{n} -{\vector{d}}_{i} )$ 
are the external vector potential of the uniform magnetic field ${\vector{B}}$ 
and scalar potential caused by the nucleus of atom $a_{i} $ 
that is located at ${\vector{R}}_{n} +{\vector{d}}_{i} $. Vectors ${\vector{R}}_{n} $ 
and ${\vector{d}}_{i} $ denote the translation vector of the lattice and vector 
specifying the position of atom $a_{i} $, respectively. 
In Eq. (\ref{eq1}), $c$, $e$ and $m$ denote the velocity of 
light, elementary charge and rest mass of electrons, respectively, and the 
matrixes $\vector{\alpha} =(\alpha_{x} ,\,\alpha_{y} ,\,\alpha_{z} )$ 
and $\beta $ stand for the usual $4\times 4$ matrices. The vector ${\vector{k}}$ 
is the wave vector that belongs to the magnetic first Brillouin zone 
\cite{11,12}. In the MFRTB method, the wave function $\Phi_{{\vectorsc{k}}} ({\vector{r}})$ 
is expanded by means of relativistic atomic orbitals for atoms 
immersed in the uniform magnetic field: 
\begin{equation}
\label{eq2}
\Phi_{{\vectorsc{k}}} ({\vector{r}})\,=\,\sum\limits_\xi {\sum\limits_n 
{\sum\limits_i {C_{{\vectorsc{k}}}^{\xi } ({\vector{R}}_{n} +{\vector{d}}_{i} )
\psi_{\xi }^{a_{i} ,{\vectorsc{R}}_{n} +{\vectorsc{d}}_{i} } ({\vector{r}})} } } ,
\end{equation}
where $C_{{\vectorsc{k}}}^{\xi } ({\vector{R}}_{n} +{\vector{d}}_{i} )$ is 
the expansion coefficient, and 
$\psi_{\xi }^{a_{i} ,{\vectorsc{R}}_{n} +{\vectorsc{d}}_{i} } ({\vector{r}})$ 
denotes the relativistic atomic orbital for 
the atom $a_{i} $ that is immersed in the uniform magnetic field. By 
neglecting both overlap integrals involving different centres and hopping 
integrals involving three different centres, matrix elements of the 
Hamiltonian are given by \cite{11}
\begin{eqnarray}
\label{eq3}
&&H_{{\vectorsc{R}}_{m} j\eta ,{\vectorsc{R}}_{n} i\xi } =\left( {\varepsilon 
_{\xi }^{a_{i} ,\,{\vectorsc{0}}} +\Delta \varepsilon_{\xi }^{a_{i} ,\,{\vectorsc{d}}_{i} } } \right)
\delta_{{\vectorsc{R}}_{m} ,{\vectorsc{R}}_{n} } 
\delta_{j,i} \delta_{\eta ,\xi } \nonumber \\ 
&+&(1-\delta_{{\vectorsc{R}}_{m} ,{\vectorsc{R}}_{n} } \delta_{j,i} 
)e^{-i{\frac{eB}{\hbar }}(R_{nx} +d_{ix} -R_{mx} -d_{jx} )(R_{my} +d_{jy} 
)\,}T_{\eta \xi }^{a_{j} a_{i} } ({\vector{R}}_{n} -{\vector{R}}_{m} +{\vector{d}}_{i} -{\vector{d}}_{j} ) 
\end{eqnarray}
with
\begin{equation}
\label{eq4}
T_{\eta \xi }^{a_{j} a_{i} } ({\vector{R}}_{l} +{\vector{d}}_{i} -{\vector{d}}_{j} )
=\int {\psi_{\eta }^{a_{j} ,{\vectorsc{0}}} ({\vector{r}})^{\dag }{\frac{v_{a_{j} } 
({\vector{r}})+v_{a_{i} } ({\vector{r}}-{\vector{R}}_{l} -{\vector{d}}_{i} +{\vector{d}}_{j} )}{2}}
\psi_{\xi }^{a_{i} ,{\vectorsc{R}}_{l} +{\vectorsc{d}}_{i} -{\vectorsc{d}}_{j} } ({\vector{r}})d^{3}r} ,
\end{equation}
\begin{equation}
\label{eq5}
\Delta \varepsilon_{\xi }^{a_{i} ,\,{\vectorsc{ d}}_{i} } 
=\int {\psi_{\xi }^{a_{i} ,\,{\vectorsc{ d}}_{i} } 
({\vector{ r}})^{\dag }
\begin{array}{l}
\left\{ 
{\displaystyle\sum\limits_{\vectorsc{{ R}}_{m}} 
\displaystyle\sum\limits_{k } 
{v_{a_{k} } ({\vector{ r}}-{\vector{ R}}_{m} -{\vector{ d}}_{k} )} } 
\right\} \\[-3mm] 
\,\,\,(\vectorsc{{ R}}_{m} \!\!+\!\!{\vectorsc{ d}}_{k} \! \ne  \!{\vectorsc{ d}}_{i} ) 
\end{array}
\psi_{\xi }^{a_{i} ,\,{\vectorsc{ d}}_{i} } ({\vector{ r}})d^{3}r} .
\end{equation}
where $T_{\eta \xi }^{a_{j} a_{i} } ({\vector{R}}_{l} +{\vector{d}}_{i} 
-{\vector{d}}_{j} )$, $\varepsilon_{\xi }^{a_{i} ,\,{\vectorsc{0}}} $ and 
$\Delta \varepsilon_{\xi }^{a_{i} ,\,{\vectorsc{d}}_{i} } $ denote the 
magnetic hopping integral, atomic spectrum and energy of the crystal field 
for the nonzero magnetic field case, respectively. In order to calculate 
$T_{\eta \xi }^{a_{j} a_{i} } ({\vector{R}}_{l} +{\vector{d}}_{i} -{\vector{d}}_{j} )$, 
$\varepsilon_{\xi }^{a_{i} ,\,{\vectorsc{0}}} $ and $\Delta 
\varepsilon_{\xi }^{a_{i} ,\,{\vectorsc{d}}_{i} } $, the perturbation theory 
is employed in the MFRTB method \cite{11}. This enables us to approximately 
express $T_{\eta \xi }^{a_{j} a_{i} } ({\vector{R}}_{l} +{\vector{d}}_{i} 
-{\vector{d}}_{j} )$, $\varepsilon_{\xi }^{a_{i} ,\,{\vectorsc{0}}} $ and 
$\Delta \varepsilon_{\xi }^{a_{i} ,\,{\vectorsc{d}}_{i} } $ by using the 
hopping integral, atomic spectrum and energy of the crystal field for the 
zero magnetic field case \cite{11}. The resultant matrix elements in the MFRTB 
method is given by 
\begin{eqnarray}
\label{eq6}
&&\!\!\!\!\!\!\!\!\!H_{{\vectorsc{R}}_{m} j({n}'{l}'{J}'{M}'),{\vectorsc{R}}_{n} i(nlJM)} \!=\!
\left(\! 
{\bar{{\varepsilon }}_{nlJ}^{a_{i} } (\!{\vector{B}}\!\!=\!\!0) \!+\! 
\Delta \bar{{\varepsilon }}_{nlJM}^{a_{i} ,\,{\vectorsc{d}}_{i} } 
(\!{\vector{B}}\!\!=\!\!0)
\!+\!
{\frac{eB}{2m}}{\frac{2J\!+\!1}{2l\!+\!1}}\hbar M}\!\! \right)
\delta_{{\vectorsc{R}}_{m} ,{\vectorsc{R}}_{n} } 
\delta_{j,i} 
\delta _{{n}'{l}'{J}'{M}',\,nlJM} \nonumber \\ 
&\!\!\!\!\!\!\!\!+\!\!&\!\!
(1\!-\!\delta_{{\vectorsc{R}}_{m} ,{\vectorsc{R}}_{n} } \delta_{j,i} )
e^{-i{\frac{eB}{2\hbar }}(R_{nx} +d_{ix} -R_{mx} -d_{jx} )(R_{ny} +d_{iy} 
+R_{my} +d_{jy} )\,}t_{{n}'{l}'{J}'{M}',\,nlJM}^{a_{j} a_{i} } 
({\vector{R}}_{n} \!-\!{\vector{R}}_{m} \!+\!{\vector{d}}_{i} \!-\!{\vector{d}}_{j} ),
\end{eqnarray}
where $t_{{n}'{l}'{J}'{M}',\,nlJM}^{a_{j} a_{i} } 
({\vector{R}}_{n} -{\vector{R}}_{m} +{\vector{d}}_{i} -{\vector{d}}_{j} )$ denotes the 
relativistic hopping integral for the zero magnetic field case, and are 
calculated by using the relativistic version of the Slater-Koster table 
\cite{11}. In Eq. (\ref{eq6}), $\bar{{\varepsilon }}_{n\ell J}^{a_{i} } ({\vector{B}}\!=\!0)$ 
and $\Delta \bar{{\varepsilon }}_{nlJM}^{a_{i} ,\,{\vectorsc{d}}_{i}} ({\vector{B}}\!=\!0)$ 
represent the energy spectrum and energy of the crystal 
field for the zero magnetic field case, respectively. The suffices $n$, 
$\ell $, $J$ and $M$ are the principal, orbital, total angular momentum and 
magnetic quantum numbers, respectively. 

In this paper, we apply Eq. (\ref{eq6}) to the simple cubic lattice immersed in the 
magnetic field, and suppose that each lattice point has one atom with one 
$s$-electron, i.e., $(n,\,l,\,J,\,M)=(n,0,1/2,\pm 1/2)$. Taking only the hopping 
integrals between nearest neighbour atoms into consideration, and using Eq. 
(\ref{eq6}), we have the simultaneous equations for expansion coefficients as 
follows \cite{12}:
\begin{eqnarray}
\label{eq7}
&& \left[ {\bar{{\varepsilon }}_{n0{1 \over 2}}^{a_{i} } ({\vector{B}}\!=\!0)
+\Delta \bar{{\varepsilon }}_{n0{1 \over 2}M}^{a_{i} ,\,0} 
({\vector{B}}\!=\!0)+{\frac{eB}{m}}\hbar M} \right. \nonumber \\ 
&& \left. {+2K_{1} \left( {n0{\frac{1}{2}},n0{\frac{1}{2}}} 
\right)_{{\frac{1}{2}}} \left\{ {\left. {\cos (2\pi k_{z} )+\cos \left( 
{2\pi \left( {k_{x} +I{\frac{p}{q}}} \right)} \right)} \right\} } \right.} 
\right]C_{{\vectorsc{k}}}^{n0{\frac{1}{2}}M} (Ia{\vector{e}}_{y} ) \\ 
&& +K_{1} \left( {n0{\frac{1}{2}},n0{\frac{1}{2}}} \right)_{{\frac{1}{2}}} 
\left[ {C_{{\vectorsc{k}}}^{n0{\frac{1}{2}}M} \left( {(I+1)a{\vector{e}}_{y} } \right)
+C_{{\vectorsc{k}}}^{n0{\frac{1}{2}}M} \left( {(I-1)a{\vector{e}}_{y} } \right)} \right]
=E({\vector{k}})C_{\vectorsc{k}}^{n0{\frac{1}{2}}M} (Ia{\vector{e}}_{y} ), \nonumber
\end{eqnarray}
with
\begin{equation}
\label{eq8}
C_{{\vectorsc{k}}}^{{n}'0{1 \over 2}{M}'} \left( {({I}'+1)a{\vector{e}}_{y} } \right)
=\left\{ {{\begin{array}{*{20}c}
 {e^{-2\pi k_{y} }C_{{\vectorsc{k}}}^{{n}'0{1 \over 2}{M}'} \left( 
0 \right)\,\,\,\,\,\,\,\,\,\,\,\,\,\mbox{for}\,\,{I}'=q-1} \hfill \\
 {C_{{\vectorsc{k}}}^{{n}'0{1 \over 2}{M}'} \left( {({I}'+1)a{\vector{e}}_{y} } \right)
\,\,\,\,\,\,\,\,\mbox{for}\,\,{I}'\ne q-1,} \hfill \\
\end{array} }} \right.
\end{equation}
\begin{equation}
\label{eq9}
C_{{\vectorsc{k}}}^{{n}'0{1 \over 2}{M}'} \left( {({I}'-1)a{\vector{e}}_{y} } \right)
=\left\{ {{\begin{array}{*{20}c}
 {e^{2\pi ik_{y} }C_{{\vectorsc{k}}}^{{n}'0{1 \over 2}{M}'} \left( 
{(q-1)a{\vector{e}}_{y} } \right)\,\,\,\mbox{for}\,\,{I}'=0} \hfill \\
 {C_{{\vectorsc{k}}}^{{n}'0{1 \over 2}{M}'} \left( {({I}'-1)a{\vector{e}}_{y} } \right)
\,\,\,\,\,\,\,\,\,\,\,\,\,\,\mbox{for}\,\,{I}'\ne 0,} 
\hfill \\
\end{array} }} \right.
\end{equation}
where $K_{1} \left( {n0{\frac{1}{2}},n0{\frac{1}{2}}} 
\right)_{{\frac{1}{2}}} $ denotes the relativistic TB parameter \cite{11}. In the 
derivation of Eqs. (\ref{eq7}) -- (\ref{eq9}), the magnitude of the magnetic field is 
assumed to be given by
\begin{equation}
\label{eq10}
B={\frac{h}{ea^{2}}}{\frac{p}{q}},
\end{equation}
where $p$ and $q$ are relatively prime integers, and $a$ denotes the lattice constant 
\cite{11,12,17}. By solving the simultaneous equations, we have $2q$ energy 
eigenvalues for each ${\vector{k}}$, and obtain $E-{\vector{k}}$ curves 
for the system immersed in the magnetic field. Since the electronic 
structure strongly depends on the rational number $p/q$ \cite{11}, we shall 
always express the magnitude of the magnetic field in terms of $p/q$. In 
the present calculations that will be shown later, we use the following 
values for the lattice constant and relativistic TB parameters:
\begin{equation}
\label{eq11}
\begin{array}{l}
 a=0.543\,(\mbox{nm}), \\ 
 \bar{{\varepsilon }}_{n0{1 \over 2}} ({\vector{B}}\!=\!0)+
\Delta \bar{{\varepsilon }}_{n0{1 \over 2}M} ({\vector{B}}\!=\!0)
=-12.1538\,\,(\mbox{eV}), \\ 
 K_{1} \left( {n0{\frac{1}{2}},n0{\frac{1}{2}}} \right)_{{\frac{1}{2}}} 
=-1.7391\,\,(\mbox{eV}). \\ 
 \end{array}
\end{equation}
These values are the same as those used in the previous works \cite{11,12}.

At the end of this subsection, let us explain in more details the reason why 
we apply the MFRTB method to the ``hypothetic'' simple cubic lattice with $s$-electrons 
instead of real materials.  
As mentioned in Sec. I, we intend to investigate unconventional oscillation phenomena 
that cannot be explained by the semiclassical approach.  
For this aim, rigorous results of the semiclassical approach such as semiclassical energy levels, 
the period and amplitude of the magnetic oscillation are indispensable.  
This is because, by investigating the discrepancy between such rigorous results of 
the semiclassical approach and the corresponding results of the MFRTB method, 
we can discuss the origin of the unconventional oscillation phenomena.  

In order to obtain such rigorous results of the semiclassical approach, we need 
the rigorous $E-\vector{k}$ curves for the zero magnetic field case.  
This is because rigorous results of the semiclassical approach are obtained from 
the extremal cross section of the Fermi surface, cyclotron effective mass and 
curvature of the Fermi surface, and these are calculated by using the $E-\vector{k}$ 
curves for the zero magnetic field case.  
For example, if the cyclotron effective mass was not rigorously calculated, then we 
could not obtain semiclassical energy levels rigorously, which causes the difficulty 
in associating the energy levels obtained by the MFRTB method with the semiclassical ones.  
Accordingly, it would be difficult to discuss the origin of unconventional oscillation phenomena 
if the cyclotron effective mass was not rigorously calculated.  
In this work, as the model system such that the $E-\vector{k}$ curves for the zero 
magnetic field case can be obtained rigorously, we adopt the simple cubic lattice system 
with $s$-electrons.  
This enables us to accurately discuss the origin of the unconventional oscillation phenomena 
that cannot be explained by the semiclassical approach.

Of course, we can obtain the $E-\vector{k}$ curves for the zero magnetic field case via the usual 
energy-band calculations such as the DFT-based energy-band calculations.  
But, such energy-band calculations contain some kinds of errors inevitably.  
Namely, errors are caused by (i) treatment of the exchange and correlation effects, 
(ii) choice of the basis function in expanding the Bloch states, 
(iii) physical meanings of the single-particle spectra, and so on. 
 As a result, errors of the $E-\vector{k}$ curves lead to those of the extremal cross section 
of the Fermi surface, cyclotron effective mass, curvature of the Fermi surface and so on, 
which become an obstacle to getting rigorous results of the semiclassical approach.

%
\subsection{Application Range of the MFRTB method}
\label{secII-B}
In the formulation of the MFRTB method, the effect of the magnetic field is 
treated as the perturbation as mentioned in the previous subsection. In this 
subsection, we discuss the application range of the resultant simultaneous 
equations (Eqs. (\ref{eq7}) -- (\ref{eq9})). 

The Dirac equation for an isolated atom, which is located at origin and is 
immersed in the uniform magnetic field, is given by 
\begin{equation}
\label{eq12}
\left( {H_{0} +{H}'} \right)\psi_{\xi }^{a_{i} ,\,{\vectorsc{0}}} 
({\vector{r}})=\varepsilon_{\xi }^{a_{i} ,\,{\vectorsc{0}}} \psi_{\xi }^{a_{i} 
,\,{\vectorsc{0}}} ({\vector{r}})
\end{equation}
with 
\begin{equation}
\label{eq13}
H_{0} =c{\vector{\alpha }}\cdot {\vector{p}}+\beta mc^{2}+v_{a_{i} } 
\left( {{\vector{r}}} \right),
\end{equation}
\begin{equation}
\label{eq14}
{H}'=ec{\vector{\alpha }}\cdot {\vector{A}}({\vector{r}}).
\end{equation}
In the MFRTB method, ${H}'$ is treated as the perturbation \cite{11}. Within the 
first-order perturbation theory, the eigenvalue $\varepsilon_{\xi }^{a_{i} 
,\,{\vectorsc{0}}} $ is approximated as $\bar{{\varepsilon }}_{nlJM}^{a_{i} } 
({\vector{B}}\!=\!0)+{\frac{eB}{2m}}{\frac{2J+1}{2l+1}}\hbar M$ \cite{11}. 
Concerning the eigenfunction, $\psi_{\xi }^{a_{i} ,\,{\vectorsc{0}}} ({\vector{r}})$ 
is approximated as the unperturbed wave function 
$\phi_{n\ell JM}^{a_{i} } ({\vector{r}})$ 
that fits on to the perturbation (zeroth-order 
wave function) \cite{11}. As a result, matrix elements of the Hamiltonian are 
given by Eq. (\ref{eq6}). 

In order to check the application range of the above-mentioned 
approximation, let us consider matrix elements of ${H}'$ with using 
eigenfunctions of $H_{0} $ ($\phi_{n\ell JM}^{a_{i} } ({\vector{r}}))$ as 
basis functions. Matrix elements of ${H}'$ are calculated as 
\begin{eqnarray}
\label{eq15}
{H}'_{{n}'{l}'{J}'{M}',\,nlJM\,} &=&\int {\phi_{{n}'{l}'{J}'{M}'}^{a_{i} } 
({\vector{r}})^{\dag }} {H}'\phi_{nlJM}^{a_{i} } ({\vector{r}}) \nonumber \\ 
&\approx&
{\frac{e}{2m}}\int {f_{{n}'{l}'{J}'{M}'}^{a_{i} } ({\vector{r}})^{\dag }{\vector{B}}\cdot ({\vector{l}}+2{\vector{s}})
f_{nlJM}^{a_{i} } ({\vector{r}})\mbox{d}^{3}r} , 
\end{eqnarray}
where $f_{nlJM}^{a_{i} } ({\vector{r}})$ is the large component of 
$\phi_{n\ell JM}^{a_{i} } ({\vector{r}})$, and is given by
\begin{equation}
\label{eq16}
f_{nlJM}^{a_{i} } ({\vector{r}})={\frac{F_{nlJ}^{a_{i} } (r)}{r}}y_{l,\,J}^{M} (\theta ,\,\phi ).
\end{equation}
Here, $y_{l,J}^{M} (\theta ,\phi )$ is the spinor spherical harmonics. In 
the derivation of Eq. (\ref{eq15}), the small component of $\phi_{n\ell JM}^{a_{i} 
} ({\vector{r}})$ is approximated by
\begin{equation}
\label{eq17}
g_{nlJM}^{a_{i} } ({\vector{r}})\approx {\frac{1}{2mc}}
{\vector{\sigma}}\cdot {\vector{p}}\,f_{nlJM}^{a_{i} } ({\vector{r}}),
\end{equation}
where $g_{nlJM}^{a_{i} } (\vector{r})$ denotes the small component of 
$\phi_{n\ell JM}^{a_{i} } ({\vector{r}})$ \cite{19}. This approximation is obtained by 
neglecting terms of order ${v^{2}}/{c^{2}}$ \cite{19}. It should be 
noted that the diamagnetic term does not appear in Eq. (\ref{eq15}) under the 
approximation Eq. (\ref{eq17}). As is well-known, the diamagnetic term is much 
smaller than the Zeeman term for the case of $B<10^{4}$ (T) \cite{20}. Therefore, if 
we did not adopt the approximation Eq. (\ref{eq17}), then additional terms that are 
related to the diamagnetic term would appear in Eq. (\ref{eq15}). Since such 
additional terms are expected to be negligibly small for $B<10^{4}$ (T), 
the approximation Eq. (\ref{eq17}) can readily be adapted except for cases of
$B>10^{4}$ (T) that corresponds to $p/q>0.713$. 

As mentioned in the previous subsection, we take $s$-electrons 
($(n,l,J,M)=(n,0,1/2,\pm 1/2))$ into consideration in the present 
calculations. Therefore, let us consider only the matrix elements that are 
related to $(n,l,J,M)=(n,0,1/2,\pm 1/2)$. From Eq. (\ref{eq15}), 
${H}'_{{n}'{l}'{J}'{M}',\,n0{\frac{1}{2}}M\,} $ ($M=\pm 1/2)$ can be calculated as
\begin{equation}
\label{eq18}
{H}'_{{n}'{l}'{J}'{M}',\,n0{\frac{1}{2}}M\,} \approx 
{\frac{eB}{m}}\hbar M\delta_{{n}',n} \delta_{{l}',0} \delta_{{J}',1/2} 
\delta_{{M}',M} ,
\end{equation}
where we again neglect terms of order ${v^{2}}/{c^{2}}$ in the 
derivation of Eq. (\ref{eq18}). 
Thus, matrixes of ${H}'$ have only diagonal elements 
with respect to $\phi_{n0{1 \over 2}M}^{a_{i} } ({\vector{r}})$ ($M=\pm 1/2)$. 
This means that both $\phi_{n0{1 \over 2}{1 \over 2}}^{a_{i} } ({\vector{r}})$ and 
$\phi_{n0{1 \over 2}-{1 \over 2}}^{a_{i} } ({\vector{r}})$ 
become eigenfunctions of not only $H_{0} $ but also $H_{0} +{H}'$ within the 
approximation of neglecting terms of order ${v^{2}}/{c^{2}}$. 
Due to Eq. (\ref{eq18}), resultant eigenvalues of $H_{0} +{H}'$ are approximately 
given by $\bar{{\varepsilon }}_{n0{1 \over 2}}^{a_{i} } ({\vector{B}}\!=\!0)
+{eB\hbar M}/m$ ($M=\pm 1/2)$. 
On the other hand, as mentioned above, 
$\phi_{n0{1 \over 2}M}^{a_{i} } ({\vector{r}})$ and 
$\bar{{\varepsilon }}_{n0{1 \over 2}}^{a_{i} } ({\vector{B}}\!=\!0)+{eB\hbar M}/m$ 
($M=\pm 1/2)$ 
are used in the MFRTB method as approximations of 
$\psi_{\xi }^{a_{i} ,{\vectorsc{0}}} ({\vector{r}})$ and $\varepsilon_{\xi }^{a_{i},{\vectorsc{0}}} $, 
respectively. Therefore, we obtain the same simultaneous 
equations as Eqs. (\ref{eq7}) -- (\ref{eq9}) if we employ the approximation of neglecting 
terms of order ${v^{2}}/{c^{2}}$ instead of using the perturbation theory. 
This means that the application range of Eqs. (\ref{eq7}) -- 
(\ref{eq9}) is not restricted by the use of the perturbation theory. Since the 
validity of the approximation of neglecting terms of order ${v^{2}}/{c^{2}}$ 
seems to be valid in the case of $B<10^{4}$ (T) ($p/q<0.713$), 
the application range of Eqs. (\ref{eq7}) -- (\ref{eq9}) 
would extend to the high magnetic field region. Therefore, we may discuss magnetic 
properties of the present $s$-electron system by means of the MFRTB method 
even for the high magnetic field case. 
%
\subsection{Boundary condition}
\label{secII-C}
In the MFRTB method, the magnetic Bloch theorem is utilized \cite{11}, which 
means that one deals with an infinitely large system. On the other hand, we 
know that the Landau-gauge vector potential that is used in the MFRTB method 
\cite{11} diverges at infinity. In order to avoid this difficulty, we introduce 
in the MFRTB method the large box, at the boundary of which a boundary 
condition is fixed appropriately. This treatment is similar to the zero 
magnetic field case, where the periodic boundary condition is utilized in 
order to treat an infinitely large system. Also in the nonzero magnetic 
field case, the boundary condition should be imposed on the wave function so 
that effects of the boundary on physical quantities make no appearance. In 
what follows, we shall explain the boundary conditions that are used in the 
actual calculations. 

The magnetic Bloch theorem is given by \cite{11}
\begin{equation}
\label{eq19}
\Phi_{{\vectorsc{k}}} ({\vector{r}}-{\vector{t}}_{n} )=
e^{i{\vectorsc{k}}\cdot {\vectorsc{t}}_{n} }e^{{ieByt_{nx} }/\hbar }
\Phi _{{\vectorsc{k}}} ({\vector{r}}),
\end{equation}
where ${\vector{t}}_{n} $ and $t_{nx} $ denote the translation vector 
defined by magnetic primitive vectors \cite{11,12} and its $x$-component. 
Concerning the wave vector ${\vector{k}}$, we have a theorem that says that 
the total number of ${\vector{k}}$ points contained in the magnetic 
first Brillouin zone coincides with that of the magnetic primitive unit 
cells contained in the system \cite{11,12}. In the case of the simple cubic 
lattice immersed in the uniform magnetic field, the magnetic primitive vectors are 
given by $a{\vector{e}}_{x} $, $qa{\vector{e}}_{y} $, and $a{\vector{e}}_{z} $. 
If we set the cube with the side length of $L=qaN$ as the large 
box, then we have $q^{2}N^{3}\,$ magnetic primitive cells in the large box 
\cite{21}. Therefore, the number of ${\vector{k}}$ points in the magnetic 
first Brillouin zone is equal to $q^{2}N^{3}$ according to the 
above-mentioned theorem. Similarly to the zero magnetic field case, we 
suppose that each component of ${\vector{k}}$ takes values with the 
interval of ${2\pi }/L$, i.e.,  
\begin{equation}
\label{eq20}
k_{x} ={\frac{2\pi }{L}}n_{x} ,\,\,\,k_{y} ={\frac{2\pi }{L}}n_{y} 
,\,\,\,k_{z} ={\frac{2\pi }{L}}n_{z} ,
\end{equation}
where $n_{x} $, $n_{y} $ and $n_{z} $ are integers. This sassumption seems 
to be reasonable because the number of ${\vector{k}}$ points in the 
magnetic first Brillouin zone is just equal to $q^{2}N^{3}$, which is 
consistent with the above-mentioned theorem. In addition, ${\vector{k}}$ 
points continuously connect to the wave vectors for the zero magnetic 
field case in the limit $B\to 0$. 

Using Eqs. (\ref{eq19}) and (\ref{eq20}), the following boundary conditions can be obtained: 
\begin{equation}
\label{eq21}
\begin{array}{l}
 \Phi_{{\vectorsc{k}}} ({\vector{r}}-L\vector{{e}}_{x} )
=e^{{ieBLy}/\hbar }\Phi_{{\vectorsc{k}}} ({\vector{r}}), \\ 
 \Phi_{{\vectorsc{k}}} ({\vector{r}}-L\vector{{e}}_{y} )
=\Phi_{{\vectorsc{k}}} ({\vector{r}}), \\ 
 \Phi_{{\vectorsc{k}}} ({\vector{r}}-L\vector{{e}}_{z} )
=\Phi_{{\vectorsc{k}}} ({\vector{r}}). \\ 
 \end{array}
\end{equation}
In the MFRTB method, we impose the boundary conditions of Eq. (\ref{eq21}) on 
$\Phi _{{\vectorsc{k}}} ({\vector{r}})$. The validity of Eq. (\ref{eq21}) is verified by 
checking the dependence of the total energy density on the size of the 
large box ($L)$. Namely, it is confirmed that the total energy density is substantially 
independent of the the size of the large box when we choose a sufficiently 
large size. Specifically, we take $200qa$ as $L$ in actual calculations \cite{12}. 
%
%
\section{Description of the de Haas-van Alphen oscillations through the 
MFTRB method}
\label{secIII}
In this section, we explain how the dHvA oscillations are described on the 
basis of the electronic structure that is calculated by the MFRTB method. 
Figure 1 shows the magnetic field dependences of the total energy and 
magnetization. The horizontal axis of Fig. 1 is $\left( {p/q} \right)^{-1}$ that 
is inversely propositional to $B$ (Eq. (\ref{eq10})). Oscillations of the total 
energy and magnetization are clearly seen in Fig. 1. In order to describe 
the oscillatory behavior, we calculate DOSs for several magnetic fields that 
are indicated by (2a) -- (2f) in Fig. 1. Resultant DOSs for the magnetic 
fields (2a) -- (2f) are shown in Figs. 2(a) -- 2(f), respectively. Peak 
positions of the DOS can be classified into two types: one is that peak 
positions increase with $p/q$, and the other is that they decrease with $p/q$. In 
Figs. 2(a) -- 2(f), peak positions of the former type are denoted by e$+$ 
and e-, and those of the latter type are denoted by h$+$ and h-, 
respectively. The pair of peaks (e$+$, e-) or (h$+$, h-) corresponds to the 
Zeeman splitting of spin states. The existence of two types of energy levels 
can be explained also by the semiclassical approach. Namely, there exist the 
electron and hole orbitals on the constant energy surface in ${\vector{ k}}$ 
space for the case of the simple cubic lattice system with $s$-electrons. 
For the present system, the cyclotron effective mass for the electron 
orbital is positive, while that for the hole orbital is negative. Since the 
interval between two energy levels is inversely proportional to the 
cyclotron effective mass according to the semiclassical approach \cite{8}, 
semiclassical energy levels that come from electron (or hole) orbitals 
increase (or decrease) with the magnetic field. Thus, we can associate e$+$ 
and e- (h$+$ and h-) with the semiclassical energy levels for electron 
(hole) orbitals. 

Next, we shall explain the relation between magnetic oscillations (Fig. 1) 
and DOSs (Figs. 2(a) -- 2(f)). It is expected that the magnetic field 
dependence of occupied energy levels near the Fermi energy has a major 
effect on that of the total energy. In the cases of the magnetic fields (2a) 
and (2b), the highest and next highest occupied energy states are e$+$ and 
e-, respectively (Figs. 2(a) and 2(b)). Therefore, the total energy is 
expected to increase with $p/q$ because their peak positions increase with $p/q$. 
Indeed, the total energy increases with $p/q$ as shown in Fig. 1. With increasing 
the magnetic field from (2b) to (2c), the highest occupied energy states 
switch from e$+$ to h-. In this situation, the highest occupied energy 
states (h-) decrease with $p/q$, while the next highest occupied energy states 
(e-) oppositely increase with $p/q$. The effect of this switch is expected to 
appear in the slope of the total energy. As shown in Fig. 1, the slope of 
the total energy is changed between (2b) and (2c). This implies that not 
only the magnetic field dependence of the highest occupied energy levels but 
also that of the next highest occupied energy levels have a major effect on 
that of the total energy. Corresponding to the change of the slope of the 
total energy, the magnetization exhibits the characteristic peak between 
(2b) and (2c) as seen in Fig. 1. When the magnetic field increases from (2c) 
to (2d), the highest occupied energy states switch from h- to e-. 
Correspondingly, the slope of the total energy slightly decreases between 
(2c) and (2d) as shown in Fig. 1. Further increase of the magnetic field 
($(2\mbox{d})\to (2\mbox{f}))$ leads to the switch of the highest occupied 
energy states from e- to h$+$. This switch results in the change of the 
slope of the total energy, which causes the kink in the magnetization 
between (2d) and (2f) (Fig. 1). At the magnetic field (2f) both the highest 
and next highest occupied energy levels decrease with $p/q$, so that the total 
energy decreases with $p/q$ (Fig. 1). 

Thus, every time the energy levels that correspond to the semiclassical 
energy level pass over the Fermi energy, the magnetic oscillation is 
produced. Namely, the dHvA oscillations are produced by the repeated 
crossing of semiclassical energy levels to the Fermi energy. This 
description of the magnetic oscillation is similar to that by 
the semiclassical approach \cite{8}.

%
\section{Additional oscillation peaks}
\label{secIV}
In this section, it is shown that additional oscillation peaks, which cannot 
be explained by the LK formula, are observed in the magnetic field 
dependence of the magnetization. Also, the origin and observability of 
additional oscillation peaks are discussed. 
%
\subsection{Additional oscillation peak and its origin}
\label{secIV-A}
Figure 3 shows the magnified view of Fig. 1. It is found from Fig. 3 that 
there exist novel and rugged peaks (additional oscillation peaks) in the 
magnetization together with the oscillations that can be explained by the LK 
formula. Of course, additional oscillation peaks cannot be explained by the 
LK formula. 

In order to clarify the origin of additional oscillation peaks, 
we calculate energy band structures for ten magnetic fields that are 
indicated by (5a) -- (5j) in Fig. 3. Figure 4 shows the energy band 
structure for the case of the magnetic field (5j), which corresponds to the 
DOS of Fig. 2(e). The horizontal axis of Fig. 4 denotes the special ${\vector{k}}$ 
points in the magnetic first Brillouin zone of the simple cubic 
lattice \cite{12}. In Fig. 4, energy bands become nearly flat between Z-point 
$(0,0,\pi/a)$ and R-point $(\pi/a,0,\pi/a)$ as well as 
between M-point $(\pi/a,\pi/{aq},0)$ and $\Gamma $-point $(0,0,0)$. 
There exist two nearly flat bands around the Fermi energy 
between Z-point and R-point. These two bands correspond to e- and e$+$ of 
Fig. 2(e) because it is confirmed that their energy levels increase with $p/q$. On 
the other hand, two nearly flat bands around the Fermi energy between 
M-point and $\Gamma $-point correspond to h- and h$+$ of Fig. 2(e). It seems 
from Fig. 4 that e-, e$+$, h- and h$+$ consist of a lot of nearly flat 
bands.  In order to clarify effects of these nearly flat bands on 
the magnetic oscillation, energy bands around the Fermi energy are shown in 
Figs. 5(a) -- 5(j) for the cases of magnetic fields (5a) -- (5j). Note that 
Fig. 5(j) is the magnified view of Fig. 4. A lot of nearly flat bands 
can be seen in Fig. 5(j) more clearly than in Fig. 4. In Fig. 5(j), the 
set of nearly flat bands above the Fermi energy corresponds to the DOS peak e- 
of Fig. 2(e), while the set of nearly flat bands below the 
Fermi energy corresponds to the DOS peak h$+$ of Fig. 2(e). 
Thus, the energy levels e- and h$+$ that correspond to the semiclassical energy levels 
consist of a lot of nearly flat bands lying within a small energy width. 
This fine energy-level structure plays a crucial role in appearance of the additional 
oscillation peaks that will be mentioned below. 

Comparing Fig. 3 with Figs. 5(a) -- 5(j), we can see that additional 
oscillation peaks are produced when energy bands that are constituents of 
the cluster cross the Fermi energy. Roughly speaking, there exist three 
blocks of energy bands in the cluster 
along both Z-R and M-$\Gamma $ lines (Figs. 5(a) -- 5(j)). 
This feature of the fine energy-level structure is 
maintained during the change of the $p/q$ ratio from 92/709 to 22/167 
as shown in Figs. 5(a) -- 5(j).  
When the first block crosses the Fermi energy with increasing the 
magnetic field from (5a) to (5c), the slope of the total energy is expected 
to change, similarly to the case of the semiclassical energy level mentioned 
in Sec. III. Indeed, the kink of the magnetization appears between (5a) and 
(5c) as shown in Fig. 3. When the second block crosses to the Fermi energy, 
the magnetization has a depressed shape between (5c) and (5f). This shape is 
also due to the change of the slope of the total energy. 
Similarly to the cases of the first and second blocks, 
the crossing of the third block to the Fermi energy 
leads to a depressed shape of the magnetization between (5f) and (5j). Thus, 
additional oscillation peaks originate from the fine energy-level structure 
that is revealed by the MFRTB method. It is should be noted that the 
crossing of only one energy band may cause the additional oscillation peak. 
If we take the step of the magnetic field more finely in the horizontal 
axis of Fig. 3, then more rugged peaks will appear in the magnetization. 

%
\subsection{Relation between the fine energy-level structure and magnetic oscillations}
\label{secIV-B}
In this subsection, the relation between the fine energy-level structure and 
magnetic oscillations is investigated in more detail. For this aim, we shall 
review the energy band structure obtained by the MFRTB method \cite{11,12}. In 
the MFRTB method, the magnitude of the magnetic field is given by Eq. (\ref{eq10}). 
As discussed in Ref. \cite{11}, the energy band structure strongly depends on the 
value of the rational number $p/q$. This is because the resultant simultaneous 
equations Eqs. (\ref{eq7}) -- (\ref{eq9}) depend on $p/q$ except for the Zeeman term 
(${eB\hbar M}/m)$ that causes the shift of the eigenvalues 
alone. Let us consider energy band structures for two magnetic fields 
$B\propto 1/{q}'$ and $B\propto p/q$, the magnitudes of 
which are nearly equal to each other, i.e., $1/{q}'\approx p/q$. As mentioned in 
Sec. II, $2{q}'$ energy bands are obtained for the case of $B\propto 1/{q}'$. 
In the case of $B\propto p/q$, we have $2q\,\,(\approx 2p{q}')$ 
energy bands that are nearly $p$ times more than that in the case of 
$B\propto 1/{q}'$. This is understood by the fact that the 
period of the translation symmetry along the $y$-direction in the case of 
$B\propto p/q$ is nearly $p$ times longer than that in the case of $B\propto 
1 /{q}'$ due to the relation ${q}'p\approx q$ \cite{11,12}. Namely, due to the 
folding of the magnetic first Brillouin zone, $p$ energy gaps may be induced 
at the boundaries of the magnetic first Brillouin zone in the case of 
$B\propto p/q$. Therefore, each energy band in $B\propto 1 /{q}'$ splits 
into $p$ energy bands, so that $2q\,\,(\approx 2{q}'p)$ energy bands appears 
in the case of $B\propto p/q$ \cite{11}. 

Since the energy bands in general overlap to each other, it is expected that 
the number of allowed bands is approximately proportional to that of energy 
bands ($2{q}')$ in the case of $B\propto 1 /{q}'$, which has been directly 
confirmed through numerical calculations \cite{11}. Namely, one allowed band 
consists of several energy bands. Let us consider again two magnetic fields 
cases: $B\propto 1 /{q}'$ and $B\propto p/q$ with $p/q\approx 1/{q}'$. 
Since the individual energy band in the case of $B\propto 1/{q}'$ splits into $p$ 
energy bands in the case of $B\propto p/q$ \cite{11}, 
an allowed band in the case of $B\propto 1/{q}'$ would split into 
multiple allowed bands in the case of $B\propto p/q$, the number of multiple 
allowed bands would be proportional to $p$. 
In the previous paper \cite{11}, we refer such multiple allowed bands 
as ``cluster''. If an allowed band in the case of $B\propto 1 /{q}'$ 
consists of $w$ energy bands, then the corresponding cluster in the 
case of $B\propto p/q$ consists of more energy bands, the number of which 
would be proportional to $wp$. 

We shall take the case of $p/q=22/167$ as an example. As mentioned in 
Sec. II, energy levels that are denoted as e$+$, e-, h$+$ and h- in Fig. 
2(e) correspond to the semiclassical energy levels. These energy levels 
correspond to nearly flat bands in the energy band structure as shown in 
Fig. 4. It is confirmed from Fig. 5(j) that the nearly flat bands of e- 
consists of 22 $(=p)$ energy bands. The same is true for e$+$, h$+$ and h-. 
This means that the semiclassical energy levels (e$+$, e-, h$+$ and h-) 
correspond to the above-mentioned cluster that contains $p$ energy bands. 
Thus, it is revealed by the MFRTB method that the semiclassical energy level 
splits into multiple energy bands that form a cluster. 

At the end of this subsection, we shall give a comment on the difference 
between the conventional dHvA oscillations and additional oscillation peaks 
on the basis of the above-mentioned knowledge about the energy band structure.  
The constituent energy bands of the cluster have the same magnetic field dependence.  
Therefore, the global dependence of the total energy 
on the magnetic field (conventional dHvA oscillations) is determined 
by the magnetic field dependence of the cluster .  The crossing of constituent 
energy bands of the cluster to the Fermi energy has a small but definite influence on 
the magnetic field dependence of the total energy, which emerges 
as the additional oscillation peaks of the magnetization.   
Consequently, we can say that the additional oscillation peaks come from the 
energy bands that forms the cluster while the conventional dHvA oscillations are 
produced by the clusters that correspond to semiclassical energy levels.

\subsection{Observability of additional oscillation peaks}
\label{secIV-C}
In this subsection, we shall discuss the observability of additional 
oscillation peaks. As mentioned in the previous section, additional 
oscillation peaks originate from energy bands that are constituents of the 
cluster. Since the energy width of the cluster (energy band width) increases 
with $p/q$ \cite{11, 22}, the splitting of energy bands in the cluster would increase 
with $p/q$. It is therefore expected that the observation of additional oscillation 
peaks becomes more feasible as $p/q$ increases. Inversely, as $p/q$ decreases, 
we need to control the value of $p/q$ (the magnitude of the magnetic field) with 
good accuracy in order to observe additional oscillation peaks. For example, 
let us consider the case where $p/q$ is equal to $3/139$ ($\simeq 2.158\times 
10^{-2})$ that is much smaller than those of cases in Figs. 5(a) -- 5(j). 
Figure 6 shows the energy band structure for this case. Although the cluster 
with 3 ($=p)$ energy bands can be clearly seen in Fig. 6, the energy width 
of the cluster is much smaller than those in cases of Figs. 5(a) -- 5(j).  
In this case, there is a possibility that additional oscillation peaks produced 
by these three energy bands is observed, if we measure the magnetic field 
dependence of the magnetization with a sufficiently fine step of the magnetic field. 

There is another case of observing additional oscillation peaks.  
As shown in Fig. 3, additional oscillation peaks appear around the magnetic 
field (5f) ($p/q=96/733)$. In the present calculations we take 
0.543 (nm) as $a$ that is equal to the lattice constant of the crystalline 
silicon, so that $p/q=96/733$ corresponds to $B=1837\,(\mbox{T})$ due 
to Eq. (\ref{eq10}). If we consider the system, the period of which is longer than 
$a=$0.543 (nm), then the magnitude of the magnetic field becomes smaller for 
$p/q=96/733$. For example, if we consider the superlattice system with the 
period $10a$, then $p/q=96/733$ corresponds to $B=18.37\,(\mbox{T})$ 
that is the experimentally available magnetic field. Thus, the additional 
oscillation peaks are measurable in the laboratory for the system with a 
long period. 
%
\section{Cyclotron effective mass}
\label{secV}
As mentioned in the previous section, the cluster that corresponds to the 
semiclassical energy level has an energy band width. It is expected that the 
energy band width may have an effect on the amplitude of the dHvA 
oscillations, because it is known that the broadening of the energy level 
leads to a reduction of the oscillation amplitude \cite{18}. According to the 
conventional LK formula, the amplitude of the dHvA oscillations depends on 
the cyclotron effective mass, curvature of the Fermi surface and relaxation 
time for scattering of electrons \cite{10}. As mentioned in Sec. I, it is 
difficult to estimate these quantities at one time from the amplitude of the 
dHvA oscillations. Before discussing the effect of the energy band width of 
the cluster on the oscillation amplitude (Sec. VI), we separately estimate 
the cyclotron effective mass alone through the MFRTB method. 

First, we explain how to estimate the cyclotron effective mass. For this 
aim, let us start with reviewing the semiclassical approach for the Bloch 
electron in the magnetic field \cite{8}. In the semiclassical approach, the 
cyclotron effective mass is defined by 
\begin{equation}
\label{eq22}
m_{c} (E,k_{z} )={\frac{\hbar^{2}}{2\pi }}{\frac{dA(E,k_{z} )}{dE}},
\end{equation}
where $A(E,k_{z} )$ is the cross sectional area of the constant energy 
surface in a plane normal to the magnetic field. According to the 
semiclassical approach, the electron goes around the edge of the cross 
section with the frequency of ${eB}/{2\pi m_{c} \left( {E,k_{z} } \right)}$. 
The quantized energy levels in the semiclassical approach satisfy 
the Bohr-Sommerfeld quantization rule 
and/or Bohr's correspondence principle \cite{8}. According to Bohr's 
correspondence principle, the difference between two adjacent energy levels 
is given by Planck's constant times the frequency of classical motion at the 
energy levels \cite{8}. Therefore, if the quantized energy level is denoted as 
$\varepsilon_{\nu } (k_{z} )$, then Bohr's correspondence principle is 
expressed by 
\begin{equation}
\label{eq-a}
\!\!\!\!\!\!\!\!\!\!\!\!\!\!\!\!\!\!\!\!\!\!\!\!\!\!\!\!\!\!\!\!\!\!\!\!\!\!\!\!
(\mbox{A})\,\,\,\,\,\, 
\varepsilon_{\nu +1} (k_{z} )-\varepsilon_{\nu } (k_{z} )
={\frac{\hbar eB}{m_{c} \left( {\varepsilon_{\nu } (k_{z} ),\,k_{z} } \right)}}. 
\end{equation}
It should be noted that Eq. (\ref{eq-a}) holds approximately for energy levels with 
very high quantum number $\nu $ \cite{8}. When we consider energy levels with 
very high quantum numbers $\nu $, $\varepsilon_{\nu +1} (k_{z} 
)-\varepsilon_{\nu } (k_{z} )$ is expected to be much less than 
$\varepsilon_{\nu +1} (k_{z} )$ and $\varepsilon_{\nu } (k_{z} )$. In this 
case, it is expected that both $m_{c} \left( {\varepsilon_{\nu +1} (k_{z} 
),\,k_{z} } \right)$ and $m_{c} \left( {\left\{ {\varepsilon_{\nu +1} 
(k_{z} )+\varepsilon_{\nu } (k_{z} )} \right\}/2,\,k_{z} } \right)$ are 
close to $m_{c} \left( {\varepsilon_{\nu } (k_{z} ),\,k_{z} } \right)$ 
because the difference $\varepsilon_{\nu +1} (k_{z} )-\varepsilon_{\nu } 
(k_{z} )$ is much less than $\varepsilon_{\nu +1} (k_{z} )$ and 
$\varepsilon_{\nu } (k_{z} )$. Therefore, we can rewrite Eq. (\ref{eq-a}) by
\begin{equation}
\label{eq-b}
\!\!\!\!\!\!\!\!\!\!\!\!\!\!\!\!\!\!\!\!\!\!\!\!\!\!\!\!\!\!\!\!\!\!\!\!
(\mbox{B})\,\,\,\,\,\,  
\varepsilon_{\nu +1} (k_{z} )-\varepsilon_{\nu } (k_{z} )={\frac{\hbar eB}{m_{c} \left( {\varepsilon_{\nu +1} (k_{z} ),\,k_{z} } \right)}},
\end{equation}
\begin{equation}
\label{eq-c}
(\mbox{C})\,\,\,\,\,\, 
\varepsilon_{\nu +1} (k_{z} )-\varepsilon_{\nu } (k_{z} )={\frac{\hbar eB}{m_{c} \left( {\left\{ {\varepsilon_{\nu +1} (k_{z} )+\varepsilon_{\nu } (k_{z} )} \right\}/2,\,k_{z} } \right)}}.
\end{equation}
The DOS obtained by the semiclassical approach has a sharp peak when the 
energy is identical with $\varepsilon_{\nu } (k_{z}^{ext} )$, where 
$k_{z}^{ext} $ denotes the wave number such that $A(E,\,k_{z} )$ has a 
extremal value, i.e., 
$\left( {{\partial A(E,\,k_{z} )}/{\partial k_{z} }} \right)_{k_{z} =k_{z}^{ext} } =0$ \cite{8}. 
This means that interval of peak positions of the DOS corresponds to 
${\hbar eB}/{m_{c}}\left({\varepsilon_{\nu } (k_{z}^{ext} ),\,k_{z}^{ext} } \right)$, 
${\hbar eB}/{m_{c}}\left({\varepsilon_{\nu +1} (k_{z}^{ext}),\,k_{z}^{ext} } \right)$ or 
${\hbar eB}/{m_{c}}\left({\left\{ {\varepsilon_{\nu +1} (k_{z}^{ext} )+
\varepsilon_{\nu } (k_{z}^{ext} )} \right\}/2,\,k_{z}^{ext} } \right)$ 
depending on the choice of the expression of 
Bohr's correspondence principle ((A), (B) or (C)). 
Therefore, we may reasonably identify interval of peak positions of 
the DOS that is calculated by the MFRTB method with 
${\hbar eB}/{m_{c}}\left({\varepsilon_{\nu } (k_{z}^{ext} ),\,k_{z}^{ext} } \right)$, 
${\hbar eB}/{m_{c}}\left({\varepsilon_{\nu +1} (k_{z}^{ext}),\,k_{z}^{ext} } \right)$ or 
${\hbar eB}/{m_{c}}\left({\left\{ {\varepsilon_{\nu +1} (k_{z}^{ext} )+
\varepsilon_{\nu } (k_{z}^{ext} )} \right\}/2,\,k_{z}^{ext} } \right)$. 

In this paper, using Eqs. (\ref{eq-a}), (\ref{eq-b}) and (\ref{eq-c}), we estimate three kinds of 
the cyclotron effective mass from intervals of peak positions of the DOS 
that is calculated by the MFRTB method. Hereafter, we denote three kinds of 
cyclotron effective mass by $m_{c}^{MFRTB(\rm{A})} (E,\,k_{z}^{ext} )$, 
$m_{c}^{MFRTB(\rm{B})} (E,\,k_{z}^{ext} )$ and $m_{c}^{MFRTB(\rm{C})} 
(E,\,k_{z}^{ext} )$, corresponding to Eqs. (\ref{eq-a}), (\ref{eq-b}) and (\ref{eq-c}). It should be 
noted that if the quantum number is so sufficiently high that Bohr's 
correspondence principle holds with good accuracy, then 
$m_{c}^{MFRTB(\rm{A})} (E,\,k_{z}^{ext} )$, $m_{c}^{MFRTB(\rm{B})} 
(E,\,k_{z}^{ext} )$ and $m_{c}^{MFRTB(\rm{C})} (E,\,k_{z}^{ext} )$ will be 
approximately equal to each other. Inversely, differences between three 
kinds of cyclotron effective masses indicate the inaccuracy of the 
semiclassical approach. 

In the case of the simple cubic lattice, $A(E,\,k_{z} )$ has extremal values 
at $k_{z}^{ext} =\pi/a$ and 0 for the electron and hole orbitals, 
respectively. Figure 7 shows energy-dependences of $m_{c}^{MFRTB(\rm{X})} 
(E,\,k_{z}^{ext} )/m$ (X$=$A, B and C) of the electron orbital ($k_{z}^{ext} 
=\pi/a)$ in the case of $p/q=1/1427\approx 6.998\times 10^{-4}$ 
($B=9.82\,(\mbox{T}))$. For comparison, the rigorous value of 
$m_{c} (E,\,k_{z}^{ext} )/m$ that is calculated from Eq. (\ref{eq22}) is also shown 
in Fig. 7 by the solid line. 
It is found that three kinds of the cyclotron effective masses 
$m_{c}^{MFRTB(\rm{X})} (E,\,k_{z}^{ext} )/m$ (X$=$A, B and C) 
are in a good agreement with $m_{c} (E,\,k_{z}^{ext} )/m$. 
Also, $m_{c}^{MFRTB(\rm{X})} 
(E,\,k_{z}^{ext} )/m$ (X$=$A, B and C) are approximately equal to each 
other. The differences between $m_{c} (E,\,k_{z}^{ext} )/m$ and 
$m_{c}^{MFRTB(\rm{X})} (E,\,k_{z}^{ext} )/m$ (X$=$A, B and C) are about 
-0.06{\%}, 0.06{\%} and -0.0002{\%}, respectively. These agreements mean 
that Bohr's correspondence principle holds with good accuracy in the case of 
$p/q=1/1427$. 

Figures 8(a) and 8(b) show the magnetic field dependences of 
$m_{c}^{MFRTB(\rm{X})} (E_{F} ,\,k_{z}^{ext} )$ for the electron orbital 
($k_{z}^{ext} =\pi /a)$ and hole orbital ($k_{z}^{ext} =0)$, respectively. 
Vertical axes of Figs. 8(a) and 8(b) denote the difference between $m_{c} 
(E_{F} ,\,k_{z}^{ext} )$ and $m_{c}^{MFRTB(\rm{X})} (E_{F} ,\,k_{z}^{ext} 
)$ ($k_{z}^{ext} =\pi /a$ and 0), which is given by
\begin{equation}
\label{eq-d}
\Delta m_{c}^{MFRTB(\rm{X})} (E_{F} ,\,k_{z}^{ext} 
)={\frac{m_{c}^{MFRTB(\rm{X})} (E_{F} ,\,k_{z}^{ext} )-m_{c} (E_{F} 
,\,k_{z}^{ext} )}{m_{c} (E_{F} ,\,k_{z}^{ext} )}}.
\end{equation}
It is found from Figs. 8(a) and 8(b) that absolute values of $\Delta 
m_{c}^{MFRTB(\rm{X})} (E_{F} ,\,k_{z}^{ext} )$ ($k_{z}^{ext} =\pi /a$ and 
0) increase with the magnetic field for three cases (X$=$A, B, C). Also, it 
is confirmed from Figs. 8(a) and 8(b) that differences between three kinds 
of cyclotron effective masses ($m_{c}^{MFRTB(\rm{X})} (E_{F} 
,\,k_{z}^{ext} )$(X$=$A, B and C)) also increase with the magnetic field. 
Since the present calculations by the MFRTB method are valid even for the 
high magnetic field region as mentioned in Sec. II, these tendencies suggest 
that accuracy of semiclassical energy levels gets worse with increasing the 
magnetic field. This can be understood by considering the maximum quantum 
number. Namely, Bohr's correspondence principle is valid for energy levels 
with very high quantum number \cite{8}. The maximum quantum number is roughly 
estimated by the ratio $E_{F} $ and ${\hbar eB}/{m_{c}\left( {E_{F},k_{z}^{ext} } \right)}$. 
This ratio becomes the order of $10^{3}$ in the case of $B\sim 
10\,(\rm{T})$ ($p/q\sim 7\times 10^{-4})$, while it is about 10 in 
the case of $B\sim 10^{3}\,(\rm{T})$ ($p/q\sim 7\times 10^{-2})$. Thus, the 
maximum quantum number increases with $B$, so that the accuracy of 
semiclassical energy levels gradually gets worse in the high magnetic field. 

It should be noted that the absolute value of $\Delta 
m_{c}^{MFRTB(\rm{C})} (E_{F} ,\,k_{z}^{ext} )$ is smaller than those of 
$\Delta m_{c}^{MFRTB(\rm{A})} (E_{F} ,\,k_{z}^{ext} )$ and $\Delta 
m_{c}^{MFRTB(\rm{B})} (E_{F} ,\,k_{z}^{ext} )$. Therefore, we had better 
use Eq. (\ref{eq-c}) instead of Eqs. (\ref{eq-a}) and (\ref{eq-b}) if we estimate the cyclotron 
effective mass from the DOS that is obtained by the MFRTB method or 
experiments such as Photoelectron Spectroscopy. 
%
\section{Analysis of the amplitude of the dHvA oscillations}
\label{secVI}

The amplitude of the dHvA oscillations is usually analyzed on the basis of 
the LK formula that includes the effect of the scattering of electrons 
\cite{5,6,7}. The effect of the scattering of electrons is incorporated into the LK 
formula by treating the quantized energy level as the broadened energy level 
with the width of $\hbar/\tau $, where $\tau $ is a relaxation 
time \cite{18}. This broadening leads to a reduction of the oscillation amplitude 
\cite{18}. In the present MFRTB method, the scattering of electrons is not taken 
into consideration. However, the cluster that corresponds to the 
semiclassical energy level looks like having an energy width as mentioned in 
Sec. IV. Therefore, it is expected that the energy width of the cluster 
will cause the reduction of the oscillation amplitude even though the 
scattering of electrons is not taken into consideration. In this section, 
the oscillation amplitude is analyzed through the MFRTB method. 
%
\subsection{Analysis method}
\label{secVI-A}
The LK formula for the total energy density at 0 (K) is given by \cite{10} 
\begin{eqnarray}
\label{eq23}
E_{total} &=&\sqrt {\frac{e^{5}}{8\pi^{7}\hbar }} 
\sum\limits_{l=1}^{} {\sum\limits_{k_{z}^{ext} } 
{{\frac{\cos \left( {\pi l \displaystyle{\frac{gm_{c} (E_{F},k_{z}^{ext} )}{2m}}} \right)
R_{D} B^{5/2}}{l^{5/2}m_{c} (E_{F},k_{z}^{ext} )\sqrt {\left| {{A}''(E_{F} ,k_{z}^{ext} )} \right|} }}
\cos \left\{ {{\frac{\hbar lA(E_{F} ,k_{z}^{ext} )}{eB}}\!-\!2\pi l\gamma 
\!+\!{\frac{\pi }{4}}} \right\}} } \nonumber \\
&+&E_{total}^{B=0} -{\frac{\chi }{2}}B^{2}
\end{eqnarray}
with 
\begin{equation}
\label{eq24}
R_{D} =\exp \left( {-2\pi^{2}{\frac{m_{c} (E_{F} ,\,k_{z}^{ext} )k_{B} 
T_{D} }{\hbar eB}}l} \right),
\end{equation}
where ${A}''(E_{F} ,k_{z}^{ext} )$ and $\gamma $ and $g$ denote the 
curvature of the Fermi surface, g-factor and phase correction, respectively. 
The factor $R_{D} $ is the so-called Dingle factor, and $T_{D} $ denotes the 
Dingle temperature that is defined by $T_{D} =\hbar/{2\pi k_{B} \tau }$ \cite{18}. 
In Eq. (\ref{eq23}), $E_{total}^{B=0} $ and $-{\chi B^{2}}/2$ denote the total 
energy density for the zero magnetic field case 
and magnetization energy density, respectively, where 
$\chi $ is the susceptibility. 

In order to analyze the amplitude of the dHvA oscillations, we determine 
values of ${A}''(E_{F} ,k_{z}^{ext} )$, $T_{D} $, $\gamma $, $A(E_{F} 
,k_{z}^{ext} )$, $E_{total}^{B=0} $ and $\chi $ by fitting Eq. (\ref{eq23}) to 
calculation results of the MFRTB method, where the value of $g$ is fixed at 
2.0 because the MFRTB method is based on the Dirac equation. The method of 
least squares is employed in the fitting procedure. As the value of $m_{c} 
(E_{F} ,\,k_{z}^{ext} )$, we use $m_{c}^{MFRTB(\rm{C})} (E_{F} 
,\,k_{z}^{ext} )$ that is evaluated in the previous section (Figs. 8(a) and 
8(b)). Specifically, the following form is employed in the fitting 
procedure:
\begin{equation}
\label{eq25}
{\frac{m_{c}^{MFRTB(\rm{C})} (E_{F} ,\,k_{z}^{ext} )-m_{c} (E_{F} 
,\,k_{z}^{ext} )}{m_{c} (E_{F} ,\,k_{z}^{ext} )}}=\mbox{3.9467}\times 
\mbox{10}^{-11}\times B^{2.7783}.
\end{equation}
This formula approximately represents both magnetic field dependences of 
$m_{c}^{MFRTB(\rm{C})} (E_{F} ,\pi /a)$ and $m_{c}^{MFRTB(\rm{C})} 
(E_{F} ,0)$ that are shown in Figs. 8(a) and 8(b), respectively.  Bearing 
in mind that both relations $A(E_{F} ,0)=A(E_{F} ,\pi /a)$ and ${A}''(E_{F} 
,0)={A}''(E_{F} ,\pi /a)$ hold for the case of the simple cubic lattice, 
parameters that should be determined in the fitting procedure are six ones, 
i.e., $A(E_{F} ,0)\,\,\left( {=A(E_{F} ,\pi /a)} \right)$, $T_{D} $, 
$A(E_{F} ,0)\,\,\left( {=A(E_{F} ,\pi /a)} \right)$, $\gamma $, 
$E_{total}^{B=0} $ and $\chi $. Values of ${A}''(E_{F} ,k_{z}^{ext} )$ and 
$T_{D} $ are related to the oscillation amplitude in a different manner, and 
those of $A(E_{F} ,k_{z}^{ext} )$ and $\gamma $ determine the period and the 
shift of the oscillation, respectively.  The non-oscillatory part of 
$E_{total} $ is determined by values of $E_{total}^{B=0} $ and $\chi $. 
Therefore, it is expected that we may readily determine these values by 
fitting Eq. (\ref{eq23}) to calculation results of the MFRTB. 
In the subsequent subsections, we discuss values of parameters that are 
related to the dHvA oscillations, i.e., $A(E_{F} ,k_{z}^{ext} )$, $T_{D} $, 
${A}''(E_{F} ,k_{z}^{ext} )$ and $\gamma $. 

%
\subsection{Low $p/q$ region}
\label{secVI-B}
As mentioned in Sec. III C, the energy width of the cluster decreases with 
decreasing $p/q$. Judging from the small energy width of the 
cluster that is obtained for the case of $p/q\simeq 2.158\times 10^{-2}$ (Fig. 
6), the energy width of the cluster would be negligible small in the low $p/q$ 
region ($p/q<<2.158\times 10^{-2})$. In addition, the 
deviation in the cyclotron effective mass is also negligible for this region 
according to the discussion of Sec. V. Therefore, it is expected that the LK 
formula works well for the low $p/q$ range. 

Values of ${A}''(E_{F} ,k_{z}^{ext} )$, $T_{D} $, $\gamma $, $A(E_{F},k_{z}^{ext} )$, 
$E_{total}^{B=0} $ and $\chi $ are determined individually for three regions of 
$p/q$ that satisfy the condition $p/q<<2.158\times 10^{-2}$: 
(a) $6.972\times 10^{-4}-7.070\times 10^{-4}$, 
(b) $3.079\times 10^{-3}-3.270\times 10^{-3}$, 
(c) $5.069\times 10^{-3}-5.666\times 10^{-3}$. 
The resultant values are summarized in Table I. Figures 9(a), 9(b) and 9(c) 
show resultant fitted curves (solid lines) and calculation results of the 
MFRTB method for three $p/q$ regions (a), (b) and (c), 
respectively. 
For reference, the magnetization that is calculated by differentiating 
the total energy curves is also shown in Figs. 9(a), 9(b) and 9(c) \cite{a1}. 
As shown in Figs. 9(a), 9(b) and 9(c), the LK formula with resultant parameters 
(Table I) well reproduces the dHvA oscillations calculated by MFRTB method. 
Oscillation periods obtained for three regions are in good agreement with 
the rigorous 
value that is calculated from the energy band structure for the 
zero magnetic field case (see, Table II). This agreement is consistent 
with the result of the previous paper \cite{12}. Values of $T_{D} $ are nearly 
equal to zero, which means that a ``pseudo'' Dingle temperature does not 
appear in these $p/q$ regions. Values of ${A}''(E_{F} ,k_{z}^{ext} 
)$ are also in good agreement with the 
rigorous value that 
is given in Table II. Differences between the 
rigorous value of 
${A}''(E_{F} ,k_{z}^{ext} )$ and fitted values are less than 0.01{\%} 
for three $p/q$ regions. 

The above-mentioned good agreements between resultant values of fitting 
parameters and rigorous values 
are consistent with the good agreement between 
$m_{c} (E_{F} ,\,k_{z}^{ext} )$ and $m_{c}^{MFRTB(\rm{X})} (E_{F} 
,\,k_{z}^{ext} )$ that is discussed in Sec. V. This suggests that the dHvA 
oscillations observed in these $p/q$ regions can be well described by the LK 
formula with good accuracy. 
%
\subsection{Hig $p/q$ region}
\label{secVI-C}
Figure 10 shows the magnetic field dependence of the total energy for the 
case of $p/q$ ranging from $2.878\times 10^{-2}$ to $5.703\times 10^{-2}$. 
In Fig. 10, plots and dashed line denote calculation results of the MFRTB method 
and those of the LK formula with rigorous values of parameters 
that are given in Table II \cite{23}. 
In this region, the energy width of the cluster would be 
non-negligible as is expected from Fig. 6 ($p/q \simeq 2.158\times 10^{-2})$. It is 
found from Fig. 10 that the total energies of the LK formula deviate from 
those of the MFRTB method with increasing $p/q$. As mentioned in Sec. II, the 
present calculations by the MFRTB method are valid even for the high $p/q$ 
region. Therefore, the deviation observed in the high $p/q$ region implies that 
the LK formula does not work well in the high $p/q$ region. 

Next, we shall discuss what kinds of errors will happen if we incorrectly 
apply the LK formula to the magnetic oscillation data for the high $p/q$ 
region. In a similar way to the previous subsection (Sec. VI. B), we 
determine parameters of the LK formula by fitting Eq. (\ref{eq23}) to calculation 
results of the MFRTB method. In the fitting procedure, the value of 
$E_{total}^{B=0} $ is fixed at the averaged value of results that are 
obtained for the low $p/q$ cases (Sec. VI B) \cite{23}. 
We determine parameters of the LK formula by the following 
two fitting procedures: 

\begin{enumerate}[(A)]
\item 
One is that $T_{D} $, $\gamma $, $A(E_{F} ,k_{z}^{ext} )$ and $\chi $ 
are used as the fitting parameters while ${A}''(E_{F} ,k_{z}^{ext} )$ is 
fixed at the rigorous value given in Table II.  
\item
Another procedure is that ${A}''(E_{F} ,k_{z}^{ext} )$, $\gamma $, 
$A(E_{F} ,k_{z}^{ext} )$ and $\chi $ are employed as the fitting parameters 
while $T_{D} $ is fixed at zero. 
\end{enumerate}
\noindent
In the former procedure the deviation of the oscillation amplitude is 
attributed to that of $T_{D} $, while it is attributed to that of 
${A}''(E_{F} ,k_{z}^{ext} )$ in the latter procedure. These fitting 
procedures are done for five $p/q$ regions: 
(a)$6.309\times 10^{-3}-8.559\times 10^{-3}$, 
(b)$8.565\times 10^{-3}-1.280\times 10^{-2}$, 
(c)$1.2848\times 10^{-2}-2.850\times 10^{-2}$, 
(d)$2.878\times 10^{-2}-5.703\times 10^{-2}$, 
(e)$5.727\times 10^{-2}-2.487\times 10^{-1}$. 

Resultant values that are determined by the fitting procedure (A) and (B) 
are summarized in Tables III and IV, respectively. It is found from Table 
III that the ``pseudo'' Dingle temperature increases with $p/q$ and reaches a 
typical order of the Dingle temperature (0.1 -- 1 (K)) that is observed in 
experiments. The reason why the pseudo Dingle temperature increases with $p/q$ is 
that the energy width of the cluster increases with $p/q$ as mentioned in Sec. 
IV C. Thus, the reduction of the oscillation amplitude is caused by the 
energy width of the cluster non-negligibly even though $\tau $ is very 
large. 

If the reduction of the oscillation amplitude is attributed to the value of 
${A}''(E_{F} ,k_{z}^{ext} )$ instead of the pseudo Dingle temperature, the 
resultant value of ${A}''(E_{F} ,k_{z}^{ext} )$ gradually increases with $p/q$ 
(Table IV). This would cause the overestimation of the curvature of the 
Fermi surface ${A}''(E_{F} ,k_{z}^{ext} )$ if the LK formula was incorrectly 
utilized in analyzing the oscillation amplitude for the high $p/q$ 
region. 

In both Tables III and IV, oscillation periods gradually increase with $p/q$, so 
that the difference between the oscillation period and 
rigorous one increases with $p/q$. This $p/q$ dependence of the period is consistent 
with the result of the previous paper \cite{12}. The difference in the period 
implies that $A(E_{F} ,k_{z}^{ext} )$ would be underestimated if the LK 
formula was incorrectly applied to the magnetic oscillation data for the 
high $p/q$ region. Although the value of $\gamma $ is close to that for the free 
electron case ($\gamma =0.5)$ in the low $p/q$ regions (Table I), it gradually 
deviates from 0.5 with increasing $p/q$ (Tables III and IV). 
This means that the free electron 
model becomes unsuitable for the system immersed 
in the high magnetic field with high $p/q$. 

It should be mentioned that the above-mentioned reduction of the oscillation 
amplitude may be observed experimentally depending on the system. As 
mentioned in Sec. IV C, the energy width of the cluster depends 
on $p/q$ \cite{11, 22}. In the case of the simple cubic lattice, it is found from Table III or 
IV that the pseudo Dingle temperature or overestimation of ${A}''(E_{F} 
,k_{z}^{ext} )$ becomes non-negligible when $p/q$ is more than $2.878\times 
10^{-2}$. The rational number $p/q\approx 2.878\times 10^{-2}$ corresponds to 400 
(T) for the system with $a=0.543$ (nm). If we consider the system with the 
period that is one order of magnitude longer than $a$, then 
$p/q\approx 2.878\times 10^{-2}$ corresponds to $B\approx 4$(T) that is 
experimentally available magnetic field. Thus, there is a possibility that 
the pseudo Dingle temperature and/or the overestimation of ${A}''(E_{F} 
,k_{z}^{ext} )$ are observed experimentally in the system with a long period 
such as a superlattice system. 

%
\section{Concluding remarks}
\label{secVIII}
The MFRTB method is the first-principles calculation method for electronic 
structures of metals immersed in the magnetic field. On the basis of 
electronic structures calculated by the MFRTB method, we investigate 
magnetic properties of the simple cubic lattice system with $s$-electrons 
that is immersed in the uniform magnetic field. The electronic structure 
calculated by the MFRTB method has the following property that becomes the 
key point for describing the magnetic oscillations of metals:

\renewcommand{\labelenumi}{(\arabic{enumi})}
\begin{enumerate}
\item The electronic structure calculated by the MFRTB method has a fine energy-level structure: The cluster of energy bands that lie within a small energy width corresponds to the semiclassical energy level. 
\end{enumerate}
\noindent
With the aid of this knowledge, we obtain the description for the 
conventional dHvA oscillations:

\begin{enumerate}
\setcounter{enumi}{1}
\item Every time the cluster of energy bands that corresponds to the semiclassical energy level crosses the Fermi energy, the slope of the total energy with respect to the magnetic field is changed, which causes the periodic change of the magnetization. 
\end{enumerate}
\noindent
The fine energy-level structure that is found by the MFRTB method causes the 
following novel phenomena:

\begin{enumerate}
\setcounter{enumi}{2}
\item When energy bands that are constituent of the cluster cross the Fermi 
energy, additional oscillation peaks of the magnetization emerge together 
with the conventional dHvA oscillations. 

\item Due to the energy width of the cluster, the unexpected reduction of the 
oscillation amplitude occurs. This reduction causes the pseudo Dingle 
temperature and/or the overestimation of the curvature of the Fermi surface. 
\end{enumerate}
\noindent
We also discuss the observability of phenomena (3) and (4), and we achieve 
the following result: 

\begin{enumerate}
\setcounter{enumi}{4}
\item There is a possibility that the above-mentioned phenomena (3) and (4) 
are observed in experiments. For example, phenomena (3) and (4) may be 
observed in some system with a long period such as a superlattice system. 
\end{enumerate}
\noindent
The MFRTB method also suggests that the semiclassical approach of the Bloch 
electron immersed in the magnetic field gets worse with increasing the 
magnetic field. Specifically, we have the following result:

\begin{enumerate}
\setcounter{enumi}{5}
\item Both the cyclotron effective mass and the period of the dHvA 
oscillations deviate from their rigorous values in the high 
magnetic field (high $p/q)$ region. These deviations would be caused by 
the fact that the highest quantum number is not as high as the semiclassical 
approximation works well. 
\end{enumerate}

Thus, beyond the semiclassical approach of the Bloch electron immersed in 
the magnetic field, the MFRTB method provides a first-principles way to 
describe physical phenomena observed in the magnetic field. Especially, the 
MFRTB method can predict the physical phenomena (such as (3) and (4)) that 
cannot be described by the semiclassical approach. 

The present work provides a novel scenario of magnetic oscillations, which will be 
effectively used when we venture into the world of real materials.  
When we apply the MFRTB method to real materials, unconventional oscillation phenomena 
such as additional oscillation peaks and unexpected reduction of the magnetic oscillation 
amplitude will emerge in the calculation results of the MFRTB method, similarly to the present case.  
If we had no knowledge about the origin of additional oscillation peaks, 
then we might incorrectly judge that the additional (non-being) cross-section of the Fermi surface 
exists because we do not know the rigorous Fermi surface for real materials.  
Also, if we had no knowledge about the origin of unexpected reduction of the 
magnetic oscillation amplitude, then we might incorrectly attribute the reduction of 
the amplitude to the cyclotron effective mass and/or curvature of the Fermi surface, 
because we do not know the rigorous values of the cyclotron effective mass and curvature 
of the Fermi surface for real materials.  
But, due to the present knowledge about origins of additional oscillation peaks and 
unexpected reduction of the magnetic oscillation amplitude, 
we will say that additional oscillation peaks may come from the fine energy-level structure 
in the case of real materials.  
Also, we will say that the unexpected reduction of the magnetic oscillation amplitude 
may originate from the energy-band width of the cluster in the case of real materials.  
Thus, the present work is indispensable for accurately discussing the origin of magnetic 
oscillations of real materials.  

In addition to the above-mentioned issue, the MFRTB method could be employed 
in solving the Kohn-Sham (KS) equation of the current-density functional 
theory (CDFT) \cite{24,25,26,27,28,29,30}. The KS equation of the CDFT contains not only the 
external vector potential but also the exchange-correlation vector potential 
that always produces a non-uniform magnetic field \cite{24,25,26,27,28,29,30}. 
In this case, relativistic atomic orbilats for the atom immersed in the non-uniform 
magnetic field ${\vector{B}}({\vector{r}})$ may be used as the basis 
functions in the expansion Eq. (\ref{eq2}). Such atomic orbitals would be 
approximated by those for the atom immersed in the uniform magnetic field 
${\vector{B}}({\vector{R}}_{n} +{\vector{d}}_{i} )$, where ${\vector{R}}_{n} +{\vector{d}}_{i} $ 
denotes the position of the atom. This 
approximation would enable us to use the perturbation theory in estimating 
the magnetic hopping integrals (Eq. (\ref{eq4})) similarly to the present MFRTB 
method \cite{11,12}. In this way, the MFRTB method will contribute to the further 
development of the first-principles way to describe physical phenomena 
observed in the magnetic field. 
\begin{acknowledgments}
This work was partially supported by Grant-in-Aid for Scientific Research 
(No. 26400354, No. 26400397 and No. 16H00916) of Japan Society for the Promotion of 
Science. 
\end{acknowledgments}
\clearpage
\begin{table}[htbp]
\caption{\label{tab1}
Resultant values of parameters for 
low $p/q$ regions.} 
\begin{center}
\begin{tabular}{p{80pt}cccc}
\hline
Range of $p/q$& 
\,\,Period($10^{\!-4}$\!/\!T)& 
$\gamma $& 
${A}''(E_{F} ,k_{z}^{ext} )$& 
$T_{\mathrm{D}}$(K)\,\,\\
\hline
$3.079\times10^{-3}$\\[-3mm]-- $3.270\times10^{-3}$& 
3.85826& 
0.500\,\, & 
8.6253430& 
$5.38\!\times\!\!10^{-7}$\\
$3.079\times10^{-3}$\\[-3mm]-- $3.270\times10^{-3}$& 
3.85855& 
0.495 & 
8.6260636& 
$7.61\!\times\!\!10^{-7}$\\
$5.069\times10^{-3}$\\[-3mm]-- $5.666\times10^{-3}$& 
3.85827& 
0.499 & 
8.6260636& 
$8.63\!\times\!\!10^{-7}$\\
\hline
\end{tabular}
\label{tab1}
\end{center}
\end{table}

\begin{table}[htbp]
\caption{\label{tab2}
Rigorous values of parameters in the LK formula.  
Rigorous values are calculated by using the energy band structure 
for zero magnetic field case. }
\begin{center}
\begin{tabular}{ccccc}
\hline
\,\,$A(E_{F} ,k_{z}^{ext} )$ (m$^{\mathrm{-2}})$& 
\,\,\,\,Period($10^{\!-4}$\!/\!T)& 
\,\,\,\,${A}''(E_{F} ,k_{z}^{ext} )$& 
\,\,\,\,\,\,$T_{\mathrm{D\thinspace }}$(K)& 
\,\,\,\,\,\,$m_{c} (E_{F} ,k_{z}^{ext} )/m$\,\, \\
\hline
$2.47411\times 10^{15}$& 
3.85826& 
8.6260636& 
0& 
0.10201 \\
\hline
\end{tabular}
\label{tab2}
\end{center}
\end{table}

\begin{table}[htbp]
\caption{\label{tab3}
Values of parameters for high $p/q$ regions. 
These values are determined by the fitting procedure (A).  }
\begin{center}
\begin{tabular}{ccccc}
\hline
Range of $p$/$q$& 
\,\,Period($10^{\!-4}$\!/\!T)\,\,&
$\gamma $& 
\,\,${A}''(E_{F} ,k_{z}^{ext} )$\,\,& 
$T_{\mathrm{D\thinspace }}$(K)\\
\hline
$6.309\times10^{-3}$\\[-3mm]-- $8.559\times10^{-3}$& 
3.85845& 
0.498 & 
8.6260636& 
$1.22\!\times\!\!10^{-4}$\\
$8.565\times10^{-3}$\\[-3mm]-- $1.280\times10^{-2}$& 
3.85870& 
0.496 & 
8.6260636& 
$1.48\!\times\!\!10^{-2}$\\
$1.2848\times10^{-2}$\\[-3mm]-- $2.850\times10^{-2}$& 
3.86005& 
0.492 & 
8.6260636& 
$1.32\!\times\!\!10^{-1}$\\
$2.878\times10^{-2}$\\[-3mm]-- $5.703\times10^{-2}$& 
3.86642& 
0.483 & 
8.6260636& 
$1.05\!\times\!\!10^{0}$\\
$5.727\times10^{-2}$\\[-3mm]-- $2.487\times10^{-1}$& 
3.88140& 
0.469 & 
8.6260636& 
$3.66\!\times\!\!10^{1}$\\
\hline
\end{tabular}
\label{tab3}
\end{center}
\end{table}

\begin{table}[htbp]
\caption{\label{tab4}
Values of parameters for high $p/q$ regions.  
These values are determined by the fitting procedure (B).  }
\begin{center}
\begin{tabular}{ccccc}
\hline
\,\,Range of $p$/$q$& 
\,\,\,\,Period (10$^{\mathrm{-4}}$/T)& 
$\gamma $& 
\,\,${A}''(E_{F} ,k_{z}^{ext} )$\,\,& 
$T_{\mathrm{D\thinspace }}$(K)\\
\hline
$6.309\times10^{-3}$\\[-3mm]--$8.559\times10^{-3}$& 
3.85844& 
\,\,\,\,0.498 \,\,\,\,& 
\,\,8.6263427\,\,\,\,& 
0\\
$8.565\times10^{-3}$\\[-3mm]--$1.280\times10^{-2}$& 
3.85870& 
\,\,\,\,0.496 \,\,\,\,& 
8.6287488& 
0\\
$1.2848\times10^{-2}$\\[-3mm]--$2.850\times10^{-2}$& 
3.86004& 
\,\,\,\,0.492 \,\,\,\,& 
8.6397337& 
0\\
$2.878\times10^{-2}$\\[-3mm]--$5.703\times10^{-2}$& 
3.86655& 
\,\,\,\,0.483 \,\,\,\,& 
8.6724848& 
0\\
$5.727\times10^{-2}$\\[-3mm]--$2.487\times10^{-1}$& 
3.88505& 
\,\,\,\,0.467 \,\,\,\,& 
9.0602895& 
0\\
\hline
\end{tabular}
\label{tab4}
\end{center}
\end{table}

\clearpage
\noindent
Figure captions

\noindent
Fig. 1: \\
Dependences of the total energy and magnetization on the inverse of 
the magnitude of the magnetic field ranging from $p/q=0.0594-0.178$. 
Symbols (2a), (2b), (2c), (2d), (2e) and (2f) indicate the magnetic fields, 
at which we calculate the DOSs (see Figs. 2(a) -- 2(f)). 
\\
\\
Fig. 2: \\
Magnetic field dependence of the DOSs for systems immersed in the 
high magnetic fields. (a) DOS for the system immersed in the magnetic fields 
(2a) of Fig. 1. (b) DOS for the system immersed in the magnetic fields (2b) 
of Fig. 1. (c) DOS for the system immersed in the magnetic fields (2c) of 
Fig. 1. (d) DOS for the system immersed in the magnetic fields (2d) of Fig. 
1. (e) DOS for the system immersed in the magnetic fields (2e) of Fig. 1. 
(f) DOS for the system immersed in the magnetic fields (2f) of Fig. 1. 
\\
\\
Fig. 3: \\
Dependence of the magnetization on the inverse of the magnitude of 
the magnetic field ranging from $p/q=0.130-0.132$. Symbols (5a) -- (5j) indicate the 
magnetic fields, at which we calculate the DOSs (see Figs. 5(a) -- 5(j)).  
\\
\\
Fig. 4: \\
Energy band structure for the case of the magnetic field (5j) that 
is indicated in Fig. 3. This energy band structure corresponds to the DOS of 
Fig. 2(e). Symboles Z, R, M and $\Gamma $ in the holaizonal axis denote 
special $\vector{k}$ points in the magnetic first Brillouin zone \cite{12}. 
Coordinates of special $\vector{k}$ points Z, R, M and $\Gamma $ are 
given by $(0,\,0,\,\pi/a)$, $(\pi/a,\,0,\,\pi/a)$, $(\pi/a,\,\pi/qa,\,0)$ 
and $(0,\,0,\,0)$, respectively. 
\\
\\
Fig. 5: \\
Magnetic field dependence of the energy band structure for the 
system immersed in the magnetic field. (a) In the case of the magnetic field 
(5a) of Fig. 3. (b) In the case of the magnetic field (5b) of Fig. 3. (c) In 
the case of the magnetic field (5c) of Fig. 3. (d) In the case of the 
magnetic field (5d) of Fig. 3. (e) In the case of the magnetic field (5e) of 
Fig. 3. (f) In the case of the magnetic field (5f) of Fig. 3. (g) In the 
case of the magnetic field (5g) of Fig. 3. (h) In the case of the magnetic 
field (5h) of Fig. 3. (i) In the case of the magnetic field (5i) of Fig. 3. 
(j) Energy band structure for the system immersed in the magnetic fields 
(5j) of Fig. 3. 
\\
\\
Fig. 6: \\
Energy band structure for the case of $p/q=3/139\approx 2.158\times 10^{-2}$. 
The inset is the magnified view of the cluster. 
\\
\\
Fig. 7: \\
Energy dependences of $m_{c}^{MFRTB(\rm{A})} (E,\,k_{z}^{ext} 
)/m$, $m_{c}^{MFRTB(\rm{B})} (E,\,k_{z}^{ext} )/m$, 
$m_{c}^{MFRTB(\rm{C})} (E,\,k_{z}^{ext} )/m$ and $m_{c} (E,\,k_{z}^{ext} 
)/m$ for the electron orbital ($k_{z}^{ext} =\pi/a)$ in the case of 
$p/q=1/1427\approx 6.998\times 10^{-4}\, (B=9.82\,(\rm {T})\,)$. 
\\
\\
Fig. 8: \\
Magnetic field dependences of $\Delta m_{c}^{MFRTB(\rm{X})} (E_{F} 
,\,k_{z}^{ext} )$(X$=$A, B and C) for (a) the electron orbital ($k_{z}^{ext} 
=\pi /a)$ and (b) the hole orbital ($k_{z}^{ext} =0)$. 
\\
\\
Fig. 9: \\
Magnetic field dependences of the total energy (filled circle) and 
magnetization (open square) in $p/q$ ranges: 
(a)$6.972\times 10^{-4}-7.070\times 10^{-4}$, (b)$3.079\times 
10^{-3}-3.270\times 10^{-3}$, (c)$5.069\times 10^{-3}-5.666\times 10^{-3}$. 
The solid line and plots denote resultant fitted curves of the LK formula and 
calculation results of the MFRTB method, respectively. 
\\
\\
Fig. 10: \\
Magnetic field dependence of the total energy for the cases of the 
magnetic field ranging from $p/q=2.878\times 10^{-2}$ to $5.703\times 10^{-2}$. 
The solid line and plots denote calculation results of the LK formula with 
rigorous values of 
parameters (Table II) and those of the MFRTB method, 
respectively. The inset is the magnified view of the dependence for the high 
magnetic field region.


\begin{thebibliography}{00}
\bibitem{21}
Note that the number of the magnetic primitive unit cells contained in 
the system is denoted by $\,N_{{\vector{t}}_{n} } $ in the previous 
papers \cite{11,12}.
\bibitem{22}
Since the number of the allowed bands is proportional to ${q}'$, the 
energy width of the allowed band is inversely proportional to ${q}'$ in the 
case of $B\propto 1/{q}'$. In the case of $B\propto p/q$, the 
allowed band in the case of $B\propto 1/{q}'$ splits in to multiple 
allowed bands that form the cluster. Therefore, the energy width of the 
cluster in the case of $B\propto p/q$ is roughly equal to the energy width 
of the allowed band in the case of $B\propto 1/{q}'$. Therefore, the energy 
width of the cluster is inversely proportional to ${q}'\approx q/p$.
\bibitem{a1}

The total energy per unit cell ($a^3E_{total}$) changes only in the 8th or 10th 
significant decimal (Figs. 9(a), 9(b) and 9(c)).  
Since this change in the total energy per unit cell is caused by passing of the 
cluster (semiclassical energy level) over the Fermi energy, this change results 
in the conventional dHvA oscillations that are usually measured in experiments.  
Here, note that the total energy cannot be measured in experiments, but the 
magnetization is actually measured.  
The measurable oscillation of the magnetization can be caused 
by the small change of the total energy per unit cell 
as shown in Figs. 9(a), 9(b) and 9(c).
\bibitem{23}
We use -15.6404163224(eV), 
9.3407x10$^{\mathrm{-6}}$, and 0.498 as the values 
of $E_{total}^{B=0} $, $\mu_{0} \chi $ and $\gamma$, 
respectively, which are obtained by averaging 
resultant values for three low $p/q$ regions (Sec. VI B).
\end{thebibliography}
\end{document}